\documentclass{aa}
\usepackage{graphicx}
\usepackage{txfonts}
\begin{document}
\title{The line-of-sight warp of the spiral galaxy ESO 123-G23}

\author{G. Gentile \inst{1} 
\and F. Fraternali \inst{2,3}
\and U. Klein \inst{1} 
\and P. Salucci \inst{4}
}

\offprints{G. Gentile, \email{ggentile@astro.uni-bonn.de}}
\institute{Radioastronomisches Institut der Universit\"at Bonn,
Auf dem H\"ugel 71, 53121, Bonn, Germany 
\and ASTRON, P.O. Box 2, 7990 AA Dwingeloo, The Netherlands
\and Kapteyn Astronomical Institute, Postbus 800, 9700 AV, Groningen, The Netherlands
\and  SISSA, via Beirut 4, 34013, Trieste, Italy
}
\date{Received / Accepted}

\abstract{We present 3-D modelling of the distribution and kinematics of
the neutral hydrogen in the spiral galaxy ESO 123-G23. The optical appearance 
of this galaxy is an almost perfectly edge-on disk, while the neutral hydrogen is found
to extend vertically out to about 15~kpc on either side of the galactic plane. 
The H{\small I} layer and the major features of the H{\small I} data cube
can be successfully explained by a model dominated by a strong
(about 30$^\circ$) line-of-sight warp. Other 
models were tried, including a flare model and a two-component model, but they clearly 
do not reproduce the data. 
This is the 
first unambiguous detection of a galactic warp that has the maximum deviation from the 
central plane almost along the line-of-sight.
No evidence for the presence of any 
companion galaxy is found in the H{\small I} data cube. 
Line-of-sight warps in edge-on galaxies are probably 
frequent, but escape detection as they are too weak. Moreover they may 
easily be mistaken as flares or 'thick disks'. A 3-D modelling of the H{\small I} 
layer as the one presented here is needed in order to distinguish between 
these possibilities.
 
}
\maketitle
\section{Introduction}

The outer parts of the disks of spiral galaxies are frequently warped
(Bosma \cite{bosma}, Garc\'{\i}a-Ruiz et al. \cite{garcia-ruiz}). Warps
are best seen with H{\small I} observations, since the neutral hydrogen extends 
out to much larger radii than the optical disk.
H{\small I} warps
obey some general rules, one of them being that they usually start around R$_{25}$ 
(Briggs \cite{briggs}). The measured angle between the 
inner plane and the outermost observed H{\small I} ring spans
a wide range of values, from a few degrees to a few tens in some exceptional cases.
Of the various possible 
orientations that warps can have, one is considered as ``unfavourable'', as the 
warp becomes less visible: this is when the largest deviation from the central 
plane occurs {\it along the line-of-sight}.

The origin of warps is still poorly understood. Different models
have been proposed, for instance: the precession of a disk embedded in a flattened 
dark halo, misaligned with the disk itself (e.g. Toomre \cite{toomre}), or the accretion of gas with
angular momentum different from that of the disk (e.g. Ostriker \& Binney \cite{ostriker}), 
but none is completely satisfactory.

In an ongoing project to study the dark matter distribution of spiral galaxies 
a sample was culled from the list of Persic and Salucci (\cite{persic})
selecting the best determined optical rotation curves; the selection
criteria were: symmetry of the H$\alpha$ rotation curve, high angular extent,
high H{\small I} flux and low luminosity (Gentile et al., in prep.)
The aim was to combine these optical data
with H{\small I} measurements tracing the gravitational potentials much further out. One
of these galaxies, the edge-on spiral ESO\,123-G23, turned out to look remarkable 
in its H{\small I} distribution. Indeed the H{\small I} emission extends
far away (up to about 15 kpc) from either side of the plane. 
Such a pattern can be produced by different phenomena including a flaring of the
outer disk, a thick layer of gas or a warp along the line-of-sight.
We have therefore modelled the H{\small I} layer of ESO\,123-G23 considering these possibilities.
Our analysis shows that the correct explanation is a warping of the disk 
along the line-of-sight. The other possibilities are
clearly ruled out.

\section{Observations}

\begin{figure}
  \resizebox{\hsize}{!}{\includegraphics{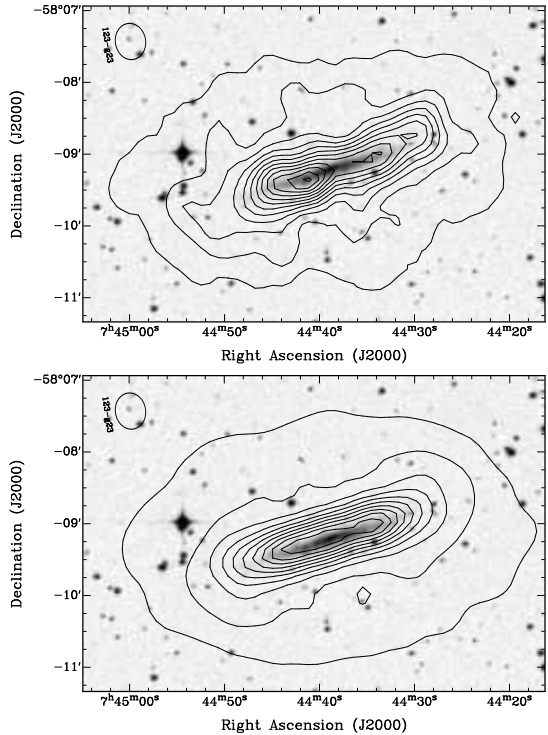}}
  \caption{Optical DSS image (grey scale) overlayed with the H{\small I} total 
intensity map of the observed data cube (top) and of the line-of-sight warp model
(bottom). Contours are 1 $\times$ $10^{20}$ cm$^{-2}$ (pseudo-3$\sigma$ 
defined similarly to Verheijen \& Sancisi \cite{verheijen}), 4, 7, 10, ... 
$\times$ $10^{20}$ cm$^{-2}$. The H{\small I} beam
is shown in the top left corner.
 }
\label{radopt}
\end{figure}

\begin{figure*}
  \resizebox{\hsize}{!}{\includegraphics{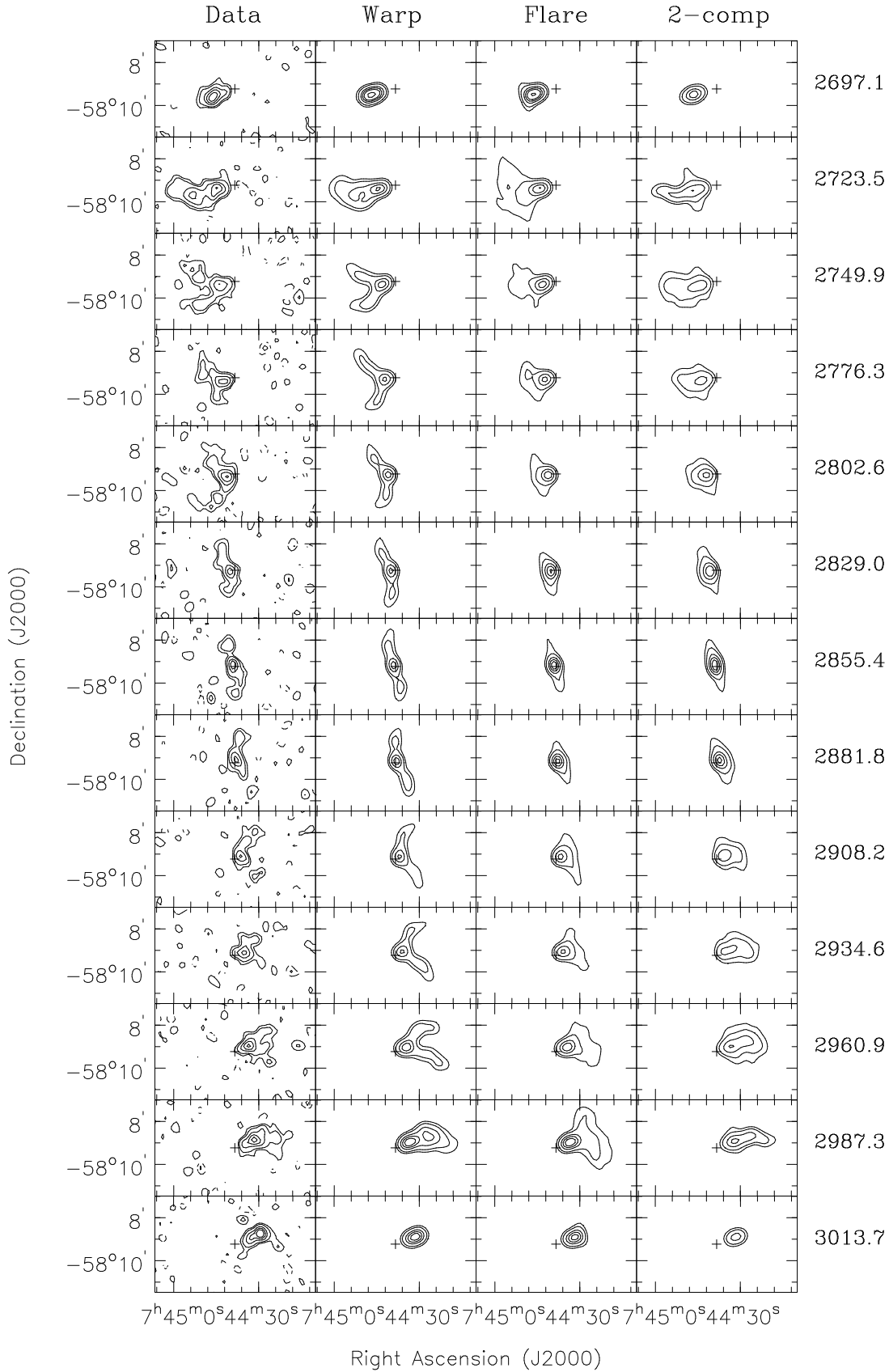}}
  \caption{
Selected channel maps; every fourth channel map is displayed. 
Contours are -4.6, -2.3, 2.3 (2~$\sigma$), 4.6, 10, 15, 20, 30, 50~mJy~beam$^{-1}$. 
The columns represent the observed data cube, the line-of-sight warp, the flare
and the two-component models. Heliocentric radial velocities are shown on right.
} 
\label{channels}
\end{figure*}

\begin{figure}
  \resizebox{\hsize}{!}{\includegraphics{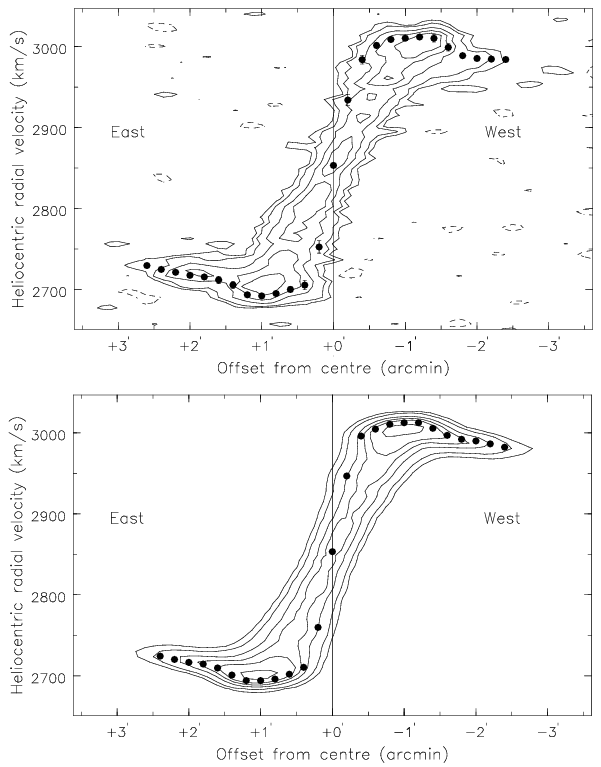}}
  \caption{
Position-velocity diagram along the central major axis of the observed data cube
(top) and of the line-of-sight warp model (bottom). Contours are
-4.6, -2.3, 2.3 (2 $\sigma$), 4.6, 10, 15, 20, 30, 50  mJy beam$^{-1}$. 
The rotation curve computed separately for the two sides is projected onto the 
top diagram and the adopted rotation curve is projected onto the bottom diagram.
The east and west sides are also indicated on the plot.
 }
\label{pvrc}
\end{figure}

ESO\,123-G23 is an edge-on spiral galaxy located at a distance of 38 Mpc 
($H_0$=75 km s$^{-1}$ Mpc $^{-1}$).  Not much was known about this galaxy 
except that it exhibits a slight warp in the outer optical 
disk (Reshetnikov \& Combes \cite{reshetnikov}). 
We have observed ESO\,123-G23 in the H{\small I} line for 2 $\times$ 12 hours 
using the Australia Telescope Compact Array
(ATCA)\footnote{The Australia Telescope is funded by the Commonwealth 
of Australia for operation as a National Facility managed by CSIRO.} and combining
the data from two array configurations (750D and 1.5A), providing a beam of 30\farcs6 
$\times$ 25\farcs0 and a velocity resolution (after Hanning smoothing) of 6.6 km s$^{-1}$. 
In Fig. \ref{radopt} (upper panel) we display an optical - H{\small I} overlay showing the peculiar 
morphology of this galaxy: while the optical image resembles an almost perfectly edge-on disk, 
the H{\small I} is much more extended perpendicularly to the major axis, 
displaying what looks like a very inclined central disk and diffuse emission at 
large (up to 15~kpc) projected distances from the optical disk;
this diffuse emission accounts for about 40 \% of the total H{\small I} mass. 
The left column of Fig. \ref{channels} shows some channel maps of the H{\small I} 
data cube.
Every fourth channel is shown, so the displayed maps are spaced by about 
26 km s$^{-1}$. The central channel map, at a velocity of 2855.4 km s$^{-1}$, is the one closer 
to the systemic velocity. The channel maps of ESO 123-G23 show the emission from a central edge-on disk, 
plus fainter emission from regions above and below the plane.
Such an emission is more visible in the central channels, close to systemic velocity.
This may indicate either a lower rotation velocity
of the gas above the plane or a lower inclination angle (i.e. towards face-on) of
the outer parts of the disk. 
A warp in position angle is visible, especially at v = 2723.5 and 
2987.3 km s$^{-1}$, and it is quite symmetric.
ESO 123-G23 is reported to have no optical companion of 
size between R$_{\rm opt}$/2 and 2 R$_{\rm opt}$ within a radius of 10 R$_{\rm opt}$ 
(Karachentsev \cite{karachentsev}). An inspection of the H{\small I} data cube did 
not disclose any companion in H{\small I} either.
The total size of the data cube is $33 \arcmin \times 33 \arcmin$ (dimensions of the
primary beam) $\times~ 442$ km s$^{-1}$. The position-velocity diagram along the
major axis is shown in Fig. \ref{pvrc} (upper panel), where the east side of 
the galaxy is slightly more extended but no clear kinematical lopsidedness is visible.
On the p-v diagram is overlapped (dots) the rotation curve of the galaxy (see Section 3).

We considered three phenomena that might mimic the observed 
H{\small I} structure and kinematics: a) a line-of-sight warp;
b) a large flare in the 
outer part of the disk; and c) a two-component structure,
thin + thick (slowly rotating) disk, 
similar to that proposed for NGC 891 (Swaters 
et al. \cite{swaters}) and NGC 2403 (Fraternali et al. \cite{fraternali}). 
These phenomena have in common the fact
of displaying emission away from the thin disk and are hard to 
disentangle except via detailed 3-D modelling of the H{\small I} layer.

\section{Models}
As mentioned above, we modelled the H{\small I} data cube
in three different ways. All these models consist of a set of concentric 
circular rings (Rogstad et al. \cite{rogstad}), each described by a set of geometrical and physical 
parameters, chosen such as to provide the
best agreement with the observed H{\small I} data cube. 
Some parameters were kept fixed for the three models. 

The parameters that the models had in common were: the radial surface density distribution,
derived using the Lucy method (Warmels \cite{warmels}); the H{\small I} velocity
dispersion, chosen to go from 12 km s$^{-1}$ in
the inner parts to 7 km s$^{-1}$ in the outer parts (Kamphuis \cite{kamphuis}),
in order to have values as realistic as possible; a constant value
of 10 km s$^{-1}$ would not have affected much the results, while values different from 10 km s$^{-1}$
by more than 5 km s$^{-1}$ would not give a good account of the observed data cube;
the shape of the vertical distribution of the H{\small I} layer, chosen to 
be Gaussian with different scale-heights for the various models 
(the choice of other distribution shapes does not 
significantly change the results); the position angle 
as a function of radius, taken to slightly change counter-clockwise after $R_{25}$
= 80\farcs7 ($\sim$15~kpc), and then to significantly (10$^\circ$) change clockwise
further out (see third panel of Fig. \ref {param}). 
By trying models with different values of the position angle we estimate an accuracy
of about 2$^\circ$ (errorbars in Fig. \ref{param}), beyond which the disagreement with
the observations is flagrant.

The rotation curve was derived with the WAMET (WArped Modified Envelope Tracing)
method (Gentile et al. \cite{gentile}, Vergani et al. \cite{vergani}), 
a modification of the Envelope Tracing 
method (Sofue \& Rubin \cite{sofue}); a detailed description of this method will 
be given in a forthcoming paper (Gentile et al., in prep.). Only small 
corrections ($<$ 5 km s$^{-1}$) were necessary to refine the rotation curve derived with our 
WAMET method. These corrections were made in order
to give a better representation of the observed H{\small I} data cube. 
This method, however, like the methods that derive the rotation curve
from the position-velocity diagram, provides the product $v_{\rm rot}~sin(i)$.
Therefore we have to account for the inclination, that can vary with radius 
within the same model. 
For the inner parts of the galaxy we used, in all the models, an inclination
angle of 85$^\circ$, a value similar to optical estimates (e.g. the Lyon-Meudon 
Extragalactic Database, LEDA, gives an inclination of 82.3$^\circ$). 

In the warp 
model the inclination angle was lowered by up to $\sim$ 30$^\circ$ beyond $R_{25}$ in order to 
fit the observations (see second panel in Fig.\ref{param}).
With the same procedure we used for the position angle, we estimated an error
for the inclination angle of 5$^\circ$. The rotation curve was obtained with the 
WAMET method with the assumed values for the inclination angle.
The FWHM of the Gaussian vertical distribution of the H{\small I} 
was chosen to be 300~pc. 

The flare model differs from the warp model in having a constant inclination
of 85$^\circ$ and a rotation curve obtained with such a constant value.
The FWHM of the
H{\small I} vertical distribution is 300~pc inside $R_{25}$ increasing to a value
of 25~kpc at $R\sim100$\arcsec. Such an unrealistic value is necessary to reproduce
the H{\small I} distribution (lower contours) visible in the total H{\small I} map 
(Fig. \ref{radopt}, upper panel). 

Finally the two-component model is the sum of a thin and a thick disk in analogy to the
previously cited models proposed for NGC\,891 and NGC\,2403. In this model the thin disk has a 
rotation velocity taken from the WAMET method with a constant inclination and
a thickness of 300 pc (FWHM). For the thick disk (40\% of the total H{\small I} mass) the rotation
velocity was lowered by 30 km s$^{-1}$ with respect to the thin disk and the thickness of the
H{\small I} layer was taken to be 20 kpc (FWHM).
Similarly to the flare model, this last unrealistic value is necessary to reproduce
the total H{\small I} distribution. For both the flare and the two-component model 
a different choice of the inclination does not improve the agreement with the
observed data cube.

\begin{figure}
  \resizebox{\hsize}{!}{\includegraphics{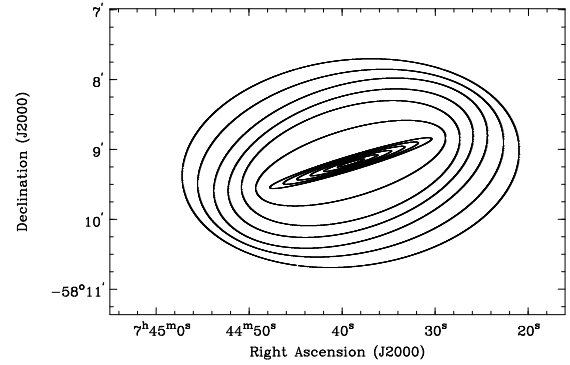}}
  \caption{
Illustration of the likely situation of the warp in ESO\,123-G23 as seen by
the observer. 
 }
\label{geom}
\end{figure}

\section{Discussion and conclusions}

In Fig.~\ref{channels} we show some channel maps of the data cube compared with the
channel maps of the line-of-sight 
warp model, the flare model, and the two-component model (from left to right). 
This figure illustrates why the flare model
and the two-component model
have to be rejected: apart from their somewhat unphysical choice of parameters 
(although necessary to produce the vertical H{\small I} distribution), they cannot
reproduce the observed channel maps. For instance, the flare 
model cannot reproduce at the same time the large north-south extension in the central 
channels and the relatively small north-south extension visible in the extreme channels
(v = 2723.5 and 2987.3 km s$^{-1}$). The 
two-component model is not able to reproduce the 'elbow' present at 
v = 2749.9 and 2802.6 km s$^{-1}$, 
which is also indicated on the receding side, at v = 2908.2 and 2960.9 km s$^{-1}$.

\begin{figure}
  \resizebox{\hsize}{!}{\includegraphics{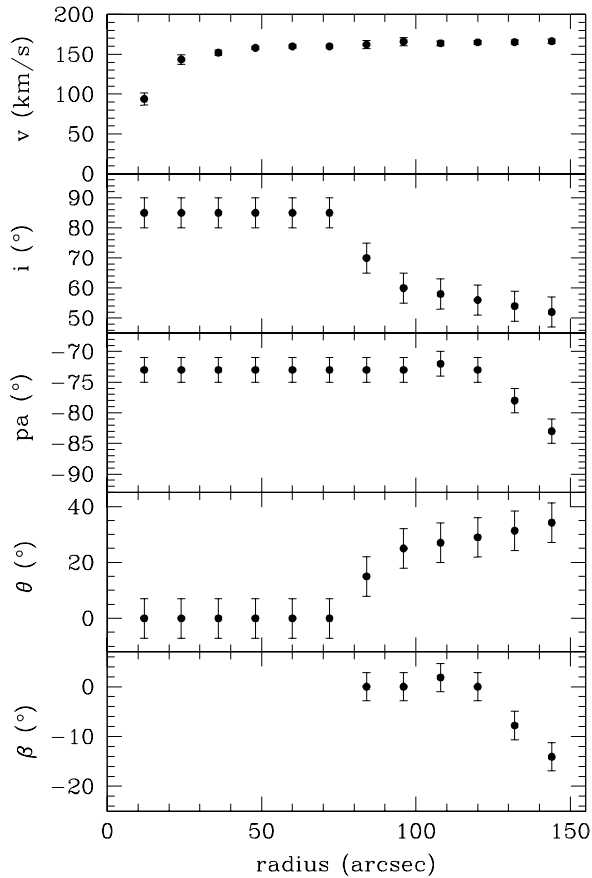}}
  \caption{
Parameters adopted for the line-of-sight warp model. From top to bottom, these 
are: rotation velocity, inclination and position angles, and the angles $\theta$ and 
$\beta$ (see text). 
 }
\label{param}
\end{figure}

The data are indeed fitted best by the model of a warp that shows maximum 
deviation from the plane close to the line-of-sight; 
Fig.~\ref{channels} shows that
such a  model reproduces in detail all the key features of the observed H{\small I} 
data cube. 
See the upper panel of Fig. \ref{param} for the adopted rotation curve
for this model; the errors are the difference between the approaching and the
receding side, considering also a minimum error of $2/sin(i)$ km s$^{-1}$.
In the upper panel of Fig. \ref{pvrc} we show the derived rotation curve of ESO 123-G23 
(separately for the two sides) projected
onto the observed position-velocity diagram,
and in the lower panel the adopted rotation curve (the average between the two sides) projected onto
the position-velocity diagram of the line-of-sight warp model.
It is remarkable that the de-warped rotation curve of this galaxy is in
good agreement with the Universal Rotation Curve of Persic et al (\cite {persic2}) 
while the raw
curve, instead, had a very unusual decrease of 30 km s$^{-1}$ from 7 kpc to 15
kpc, never found in galaxies with the maximum circular velocity of ESO 123-G23.
Fig.~\ref{geom} shows the H{\small I} rings of the
line-of-sight warp model as seen by the observer
and Fig. \ref{radopt} (lower panel) shows the total H{\small I} map obtained with
this model. The overall H{\small I} distribution is well reproduced.
We have also tried combinations of the various models, but the emission at
projected distances of up to 15 kpc from the central disk can be explained only
with the warp model, and combining the warp model with realistic values of 
an outer flare and a thick disk gives irrelevant changes to the model.

In order to describe the geometry of the warp, 
as previously done by Schwarz (\cite {schwarz}), we define the angle $\theta$ 
between the plane of the inner disk and the plane of the warped ring, and the 
angle $\beta$ between the line of nodes (at the intersection of the plane of 
the ring and the plane of the inner disk) and the line of intersection between 
the plane of the sky and the plane of the inner disk. The uniqueness of this warp
now shows up in Fig.~\ref{param}, bottom panel: $\beta$ is always relatively 
close to zero, meaning that the maximum tilt of rings is almost exactly along the 
line-of-sight. We point out that a negative value of $\beta$ means a position angle 
varying clockwise and that in unwarped sections of a galaxy $\beta$ is of course 
not defined. The fourth panel from top of Fig.~\ref{param} shows the monotonically 
increasing warp angle $\theta$ as a function of radius.

There is, however, an ambiguity due to our assumption of a central inclination of 
85$^\circ$ and the unknown true orientation of the disk: the inclination varies from 
85$^\circ$ to about 55$^\circ$, which could mean a variation of 30$^\circ$ or 40$^\circ$,
depending on the viewing angle. If we look at both the central disk and
the outer rings ``from above'' or
``from below'', the warp angle is about 30$^\circ$; if we look at one ``from above'' and the
other ``from below'', the warp angle is about 40$^\circ$.
We plot the warp angles with the assumption of viewing the galaxy in such a way 
that the warp is as weak as possible (30$^\circ$);
considering the uncertainties in position angle and inclination, we estimate
that the minimum warp angle necessary to reproduce the data cube is about 25$^\circ$. 
Even with this
assumption, this warp remains exceptional: the warp angle, measured at the 
last point, is very large. In fact it is larger for example than all the warp 
angles found in the systematic study of galaxy disks in H{\small I} by Garc\'{\i}a-Ruiz et al. 
(\cite {garcia-ruiz}). This is probably the reason why this warp is rendered 
visible: for smaller angles a warp oriented along the line-of-sight is 
almost invisible.

We have thus found very strong evidence for ESO 123-G23 to be an almost perfectly edge-on 
disk galaxy with a strong warp along the line-of-sight. This is demonstrated by the 
comparison of a series of models with the observations.
We therefore report the first {\it clear} case of a galaxy with such a warp.
We note that Becquaert \& Combes (1997) interpreted the morphology and kinematics
of the H{\small I} disk of NGC 891 in terms of a warp close to the line-of-sight. However,
the fact that Swaters et al. (1997) interpreted their H{\small I} observations of NGC 891
predominantly in terms of a thin + (slowly rotating) thick disk shows that the case was not as
evident as the present one.
Statistically, the frequency of occurrence of line-of-sight
warps should be much higher, while most of them remain undisclosed owing to
their small warp angles. The only way to discriminate between different phenomena
(warp, flare, etc.) is detailed modelling in the way described here.

\begin{acknowledgements}
We thank the referee, Rob Swaters, for valuable comments that improved the
quality of this paper.
We are grateful to Renzo Sancisi for giving us the first hints for this work.
GG wants to thank Christian Br\"uns and Daniela Vergani for performing the first
run of the observations. GG is also grateful for financial support of the {\it 
Deutsche Forschungsgemeinschaft} under number GRK 118 ``The Magellanic System, 
Galaxy Interaction and the Evolution of Dwarf Galaxies''.
\end{acknowledgements}

\end{document}